\documentclass[aps,pra,twocolumn,nopacs,floats,superscriptaddress]{revtex4-2}

\usepackage{graphicx}
\usepackage{dcolumn}
\usepackage{bm}
\usepackage{amsmath}
\usepackage{color}
\usepackage{amssymb}
\usepackage{ulem}
\usepackage{xcolor}
\usepackage{subfigure}

\usepackage{amssymb}
\usepackage{latexsym}
\usepackage{epsfig}

\usepackage{amsmath,amsfonts,amssymb,bm,ulem}
\usepackage{color}
\usepackage{float}

\definecolor{cardinal}{rgb}{0.6,0,0}
\definecolor{darkgreen}{rgb}{0,0.4,0}
\definecolor{golden}{rgb}{0.92, 0.7, 0}
\definecolor{midnight}{rgb}{0, 0, 0.5}
\definecolor{darkblue}{rgb}{0, 0, 0.7}

\def\he4{$^4$He}
\def\hel3{$^3$He}
\def\Am3{\AA$^{-3}$}
\def\beq{\begin{equation}}
\def\eeq{\end{equation}}

\newcommand{\be}{\begin{equation}}
\newcommand{\ee}{\end{equation}}
\newcommand{\bea}{\begin{eqnarray}}
\newcommand{\eea}{\end{eqnarray}}
\newcommand{\bse}{\begin{subequations}}
\newcommand{\ese}{\end{subequations}}

\begin{document}

\title{Autonomous dynamics of two-dimensional  insulating domain with superclimbing edges}

\author{Anatoly Kuklov}
\affiliation{Department of Physics \& Astronomy, College of Staten Island and the Graduate Center of
CUNY, Staten Island, NY 10314}

\author{Nikolay Prokof'ev}
\affiliation{Department of Physics, University of Massachusetts, Amherst, MA 01003, USA}

\author{Boris Svistunov}
\affiliation{Department of Physics, University of Massachusetts, Amherst, MA 01003, USA}

\begin{abstract}
Superclimbing dynamics is the signature feature of transverse quantum fluids describing wide superfluid one-dimensional interfaces and/or edges with negligible Peierls barrier. Using Lagrangian formalism, we show how the essence of the superclimb phenomenon---dynamic conjugation of the fields of the superfluid phase and geometric shape---clearly manifests itself via characteristic modes of autonomous motion of the insulating domain (``droplet") with superclimbing edges. In the translation invariant case and in the absence of supercurrent along the edge, the droplet demonstrates ballistic motion with the velocity-dependent shape and zero bulk currents. In an isotropic trapping potential, the droplet features a doubly degenerate sloshing mode.  The period of the ground-state evolution of the superfluid phase (dictating the frequency of the AC Josephson effect) is sensitive to the geometry of the droplet. The supercurrent along the edge dramatically changes the droplet dynamics: The motion acquires features resembling that of a 2D charged particle interacting with a perpendicular magnetic field. In a linear external potential (uniform force field), the state with a supercurrent demonstrates a spectacular gyroscopic effect---uniform motion in the perpendicular to the force direction.
\end{abstract}

\maketitle

\section{Introduction}

Recent progress in developing the theory of transverse quantum fluids (TQF)---quasi-one-dimensional edge superfluids featuring (at low enough temperature) stable persistent currents and off-diagonal long-range order
thanks to their infinite effective compressibility enabled by the coupling to a particle reservoir in the transverse to the superflow direction \cite{Kuklov2022,Radzihovsky2023,Kuklov2024a,Kuklov2024b,Zhang2024}---was originally
motivated by supertransport through a structurally imperfect crystal of \he4 phenomena \cite{Hallock,Hallock2012,Hallock2019,Beamish,Beamish_2016,Moses,Moses2019,Moses2020,Moses2021}. Lately,
quantum fluctuations of the edge shape were systematically addressed and demonstrated to be as interesting and
informative as superfluid properties \cite{Zhang2024}. For a review of all these activities, see Ref.~\cite{ARCMP2024}. 

The long-wave shape dynamics of the (microscopically) quantum rough superfluid edge stems from the superclimb effect---the edge motion in the direction transverse to its orientation supported by the supertranport of matter to/from the corresponding edge element \cite{sclimb}.  Quantitatively, the superclimbing dynamics is described by Hamiltonian formalism in which the field of the transverse (say, vertical) displacements of the edge is canonically conjugate to the field of the superfluid phase \cite{sclimb}. In other words, the role of the density as a conjugate variable to the phase (as explained in textbooks on superfluidity; see, {\it e.g.}, Ref.~\cite{book}) is played by the vertical displacement of the edge. At the quantum level, the ground-state and low-temperature fluctuations of the two fields are statistically coupled. This allows one to access superfluid properties of the edge by studying fluctuations of its shape \cite{Zhang2024}.

\begin{figure}[tbh]
\includegraphics[width=0.9 \columnwidth]{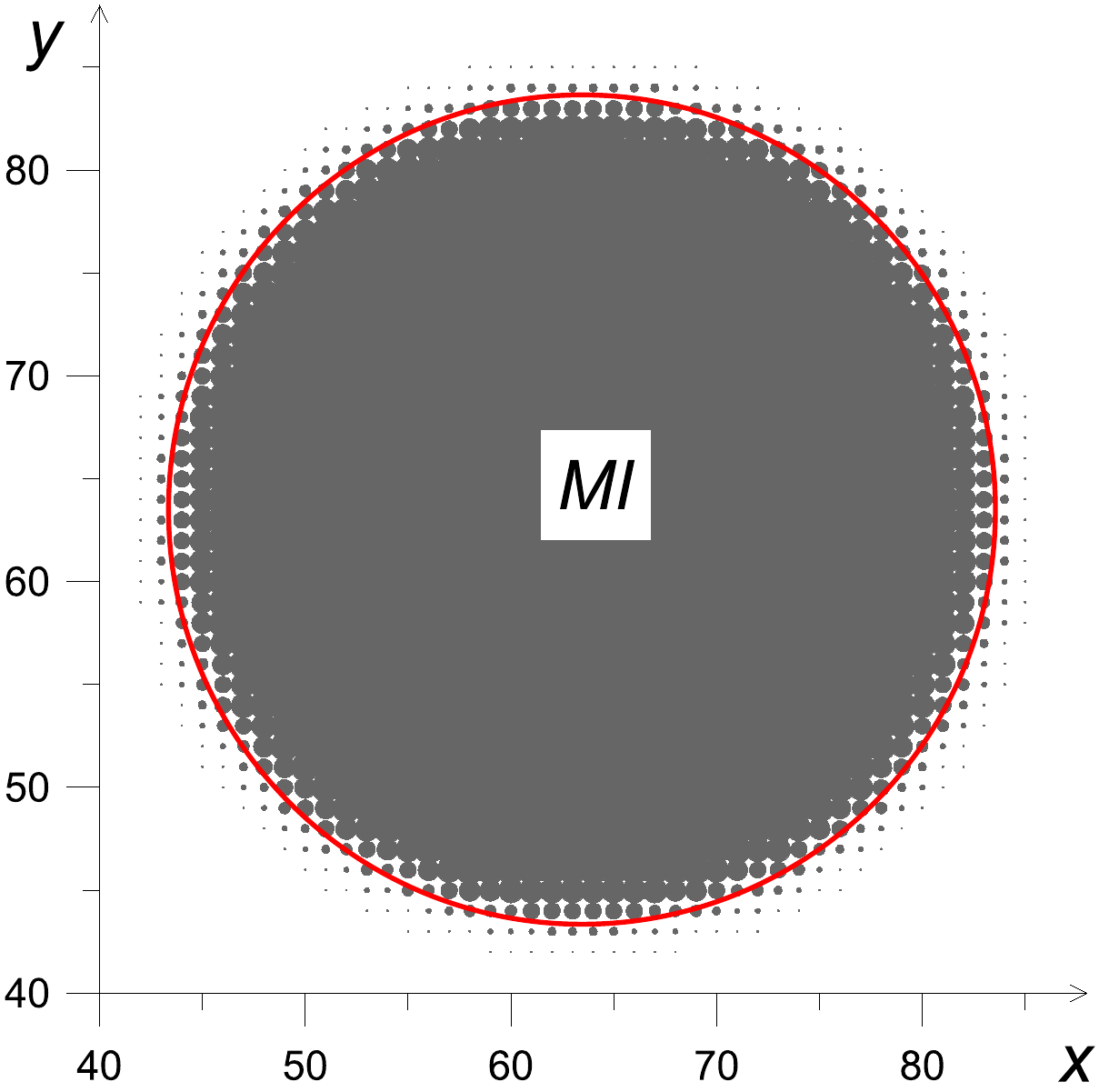}
\caption{Insulating domain with 1260 particles in model (\ref{Hhc}) at $V=-2.2t$ (symbol sizes are
proportional to the average occupation number).
Simulations were performed at temperature $T =t/128$. The red line is a circle used
for better visualization of the domain shape. 
 By symmetry, a void in the insulator occupying the outer space has the same properties. }
\label{fig1}
\end{figure}
In this work, we observe that there exists yet another---purely classical-field---way of revealing the key aspects of the superclimbing dynamics, including the ones involving persistent-current states, through dynamics of the insulating domain with superclimbing edges. Such a domain in the system of hard-core bosons with the nearest-neighbor hopping and interaction terms on the square lattice is shown in
Fig.~\ref{fig1}  (numerous alternative setups are mentioned in Sec.~\ref{sec:conclus}).  To be more specific, the self-bound domain state is described by the Hamiltonian
\begin{equation}
H_{\rm hc} = -t \sum_{\langle i,j \rangle } b_{j}^{\dagger} b_{i}^{\,}
        + V  \sum_{\langle  i,j \rangle } n_{j} n_{i}   \qquad (V<-2t)\, ,
\label{Hhc}
\end{equation}
where $b_{i}$ is the bosonic annihilation operator and occupation numbers obey
$n_{i}\le 1$. It can be re-written identically as the easy-axis spin-$1/2$ ferromagnetic
model with $J_x=J_y=2t$ and $J_z=-V$. The width of the domain edge and its superfluid stiffness are
controlled by the nearest neighbor attraction and diverge at $V \to -2t$. As demonstrated in
Ref.~\cite{Zhang2024}, when the edge width exceeds several lattice periods, the edge enters
the TQF regime when the Peierls barrier can be neglected on exponentially large scales---much larger than the domain perimeter.
Under these conditions, the discrete translation symmetry and the square-lattice discrete rotation symmetry
become irrelevant as well and the equilibrium domain shape is expected to be a perfect circle;
see Fig.~\ref{fig1}.

The emergent (upon coarse-graining) translational invariance guarantees that the droplet can perform ballistic motion with velocity-dependent shape in the absence of the bulk currents. The droplet confined by
an isotropic external potential has to demonstrate a doubly degenerate sloshing mode. In the presence of persistent current around the droplet edge, the degeneracy of the sloshing mode is lifted, and the sloshing motion is accompanied by Foucault-like precession. In a linear external potential (uniform force field), the state with a supercurrent demonstrates a spectacular gyroscopic effect---uniform motion in the perpendicular to the force direction.

\section{The model}

Quantitative analysis is conveniently performed with the Lagrangian formalism in terms of the edge position, ${\bf r}(\xi,t)=(x(\xi,t),y(\xi,t))$, and the superfluid phase along the edge, $\phi(\xi,t)$, as functions of time  $t$, with parameter $\xi$ labelling the edge points. The structure of the Lagrangian readily follows from the continuity equation, the form of the latter expressing the law of conservation of matter under specific conditions of (i) supertransport along the edge and (ii) insulating incompressible bulk.

\subsection{Parameterization freedom}

Formally, it is convenient to view the three functions $x(\xi,t), y(\xi,t), \phi(\xi,t)$ as dynamical fields despite that there should be only two conjugate variables. The redundancy in this approach  is associated with the parameterization freedom. If ${\bf r}(\xi,t)$, $\phi(\xi,t)$
is a solution to our dynamic problem, then, treating parameter $\xi$ as an {\it arbitrary} function of a new parameter, $\xi'$, and time, that is, substituting $\xi\equiv \xi(\xi',t)$ into the solution,
we get an equivalent solution, ${\bf r}'(\xi',t)$, $\phi'(\xi',t)$, where
\be
{\bf r}'(\xi',t ) = {\bf r}\left( \, \xi(\xi',t),\, t \right)\, , \quad \phi'(\xi',t ) = \phi \left( \, \xi(\xi',t),\, t \right) .
\label{reparameterization}
\ee

Thus, for the three variables there should be only two physical equations of motion, with a freedom of choosing this or that condition fixing the ``gauge,"  that is by selecting a specific parameterization.

In the vast majority of cases, one would {\it ultimately} prefer to work with two rather than three
unknown functions using parameterization of the following type:
 \be
x(\xi,t) = X(\eta(\xi,t),\xi,t) , \quad y(\xi,t) = Y(\eta(\xi,t),\xi,t) .
\label{par_gen}
\ee
Here $X(\eta,\xi,t)$ and $Y(\eta,\xi,t)$ are certain fixed functions of three variables and $\eta(\xi, t)$ is the unknown field conjugate (but not necessarily canonically) to the field $\phi(\xi, t)$. A very important practical example of parameterization (\ref{par_gen}) is the polar coordinate system with a moving origin:
 \bea
x(\theta,t) &= &x_0(t) + r(\theta,t)\cos \theta  \, , \label{polar_x}\\
y(\theta,t) &=& y_0(t) + r(\theta,t)\sin \theta  \, ,
\label{polar_y}
\eea
where the radius $r(\theta,t)$ plays the role of the function $\eta(\xi,t)$ with $\xi\equiv \theta$.

Fixing the gauge by Eqs.~(\ref{par_gen}) can be implemented at three different stages: Option~1 is to introduce the Lagrangian or Hamiltonian description directly in terms of two rather than three fields. Such an approach was used in Ref.~\cite{sclimb}, where the role of parameter $\xi$ was played by the coordinate $x$, in which case the field $y(x)$ was argued to be canonically conjugate to the field $\phi(x)$. Along similar lines, in polar coordinates, identifying parameter $\xi$ with the polar angle $\theta$,  one can see that the variable canonically conjugate to $\phi(\theta)$ is one half of the square of the polar radius, $r^2(\theta)/2$. This observation then immediately leads to the Hamiltonian of the system.

More flexible is Option~2 where one formulates a generic Lagrangian in terms of  the three fields, $x(\xi), y(\xi), \phi(\xi)$, and then substitutes Eqs.~(\ref{par_gen}) into the Lagrangian to get a gauge-specific Lagrangian in terms of  $\phi(\xi)$ and $\eta(\xi)$.  Even more flexible is Option~3, where the substitution (\ref{par_gen}) is implemented at the level of the generic equations of motion, obtained from the generic Lagrangian; or not used at all, so that the gauge is fixed by a different protocol explained in Sec.~\ref{subsec:Covariance}.

\subsection{Covariant representation}
\label{subsec:Covariance}

Gauge redundancy associated with the parameterization freedom implies that the system of three equations of motion following from the generic Lagrangian should be degenerate: One should be able to reduce it to two independent---and incomplete in view of the gauge freedom---equations and a trivial identity. From superfluid hydrodynamics it is clear that the system of two independent equations can be cast into the form when one of them is the continuity equation expressing the law of conservation of matter and the other one is the generalized Beliaev--Josephson--Anderson (BJA) equation describing the time evolution of the field $\phi$. In what follows we will see that that is indeed the case. Meanwhile, it is important to discuss the optimal representation of these equations.

We want the two dynamic equations to be maximally insensitive (covariant) with respect to the choice of parameterization. To this end, we employ the tools of differential geometry.
Let $l$ be the (algebraically understood) arc length, with its infinitesimal element being
\be
dl \, = \, l_\xi  \, d\xi \, , \qquad \quad  l_\xi \, =\, \sqrt{x^2_\xi + y_\xi^2}
\label{arclength}
\ee
($x_\xi \equiv \partial_\xi x$,  $y_\xi \equiv \partial_\xi y$, $\partial_\xi  \equiv \partial  / \partial \xi$).
The unit vectors
 \be
\hat{\bf n}\, =\,  l_\xi^{-1}(y_\xi, -x_\xi) \, , \qquad \hat{\bf t}\, =\, l_\xi^{-1} (x_\xi, y_\xi) \, ,
\label{hat_n}
\ee
are normal and tangent to the line, respectively. We will also need the signed line curvature ($x_{\xi \xi} \equiv \partial_\xi^2  x$,  $y_{\xi \xi} \equiv \partial_\xi^2 y$):
\be
\kappa \, =\,  l_\xi^{-3} (x_\xi y_{\xi \xi} - y_\xi x  _{\xi \xi} ) \, .
\label{curvature}
\ee
Finally, an important role will be played by the arc-length derivative:
\be
\partial_l \, =\,  l_\xi^{-1} \partial_\xi \, .
\label{partial_l}
\ee
The vectors  $\hat{\bf n}$ and $ \hat{\bf t}$, the scalar $\kappa $,  and the operator $\partial_l$ have purely geometric meaning rendering them covariant.
Expressing the two dynamic equations in terms of these objects leaves potentially non-covariant only the terms with time derivatives: $\dot{\bf r}$ and $\dot{\phi}$.

Considering time derivatives, our first observation is that the scalar $\hat{\bf n}\! \cdot\!  \dot{\bf r}$ is also gauge invariant. This can be shown purely mathematically but is also immediately clear from the physical meaning of this quantity---the velocity of the displacement of the edge in the vertical direction. Our second observation is that the scalar $\hat{\bf t}\! \cdot \! \dot{\bf r}$ definitely depends of the gauge and by no means can be unambiguously defined unless some extra (gauge-fixing) condition is applied. Indeed, the meaning of this quantity is the velocity of the motion of the label $\xi$ along the edge. There are absolutely no physical consequences associated with this ``motion" leaving the shape of the edge intact. But there is a mathematical consequence---a ``conspiracy" between $\hat{\bf t} \! \cdot \! \dot{\bf r}$, $\dot{\phi}$, and $\partial_l \phi$ that can be cast in the form of the invariance with respect to the gauge transformation,
\bea
\hat{\bf t}\! \cdot \! \dot{\bf r}\, &\to& \, \hat{\bf t}\! \cdot\!  \dot{\bf r} \, + \, g(\xi,t) \label{gauge1} \\
\dot{\phi}  \, &\to&\, \dot{\phi}  \, +\,  g(\xi,t) \,  \partial_l \phi \, , \label{gauge2}
\eea
where $g(\xi,t)$ is an arbitrary function;  in terms of the reparameterization (\ref{reparameterization}) it is expressed as
$g=l_\xi \dot{\xi}(\xi',t)$. The meaning of the transformation (\ref{gauge1})--(\ref{gauge2}) is purely geometrical (and, in particular, has nothing to do with the physical meaning of $\phi$). The transformation expresses the rather obvious fact that the motion of the label along the line creates an apparent contribution to time derivatives of the fields proportional to the velocity $\hat{\bf t}\! \cdot \!  \dot{\bf r}$ and the arc gradient of the field.

Hence, if all the time derivatives in the (otherwise covariant) equations of motion are expressed in terms of $\hat{\bf n}\! \cdot \! \dot{\bf r}$, $\hat{\bf t}\! \cdot \! \dot{\bf r}$, and $\dot{\phi}$, then $\dot{\phi}$ and $\hat{\bf t}\! \cdot\!  \dot{\bf r}$ have to enter the equations of motion in the form of
\be
\dot{\phi} - (\hat{\bf t} \! \cdot\! \dot{\bf r}) \,   \partial_l \phi  \qquad  \mbox{(covariant time derivative)}\, .
\label{time_covar}
\ee
Furthermore, since the time derivative of $\dot{\phi}$ is not the part of continuity equation, the term (\ref{time_covar}) should enter only the BJA equation.

Gauge-invariance relations (\ref{gauge1})--(\ref{gauge2}) suggest the following on-the-fly gauge-fixing protocol, which appears very appropriate for numeric simulations. In this protocol, one assigns any desired value, including zero, to the longitudinal velocity $\hat{\bf t} \! \cdot\! \dot{\bf r}$ at the time $t$ and parameter $\xi$. The covariant time derivative  (\ref{time_covar}) in the BJA equation then automatically assigns the matching value to 
 $\dot{\phi}$.
\subsection{Continuity equation}

The superfluid current along the edge is given by
\be
j\, =\, n_s \partial_l \phi \, ,
\label{j}
\ee
where
$n_s$ is the superfluid stiffness. As long as we are interested in the regime of an appropriately wide edge and not so large values of $j$, we can safely treat $n_s$ as a constant. (Otherwise, we would need to take into account the dependence of $n_s$ on $j$, as well as on the line orientation and curvature.)

The divergence of the superfluid current, $\partial_l j$, yields the local
(and algebraically understood ) accumulation of matter. Since the bulk is incompressibile, the edge shifts
accordingly in the transverse direction and the matter balance is expressed by the {\it continuity equation:}
\be
 \hat{\bf n}\cdot \dot{\bf r} \, +\,  \partial_l j \, =\, 0 \quad \Leftrightarrow \quad \hat{\bf n}\cdot \dot{\bf r}\, =\, -  n_s \partial_l^2 \phi \, .
\label{cont}
\ee
Here and in what follows the unit of length is defined by the condition that the 2D particle number density in the bulk equals unity.

Integrating Eq.~(\ref{cont}) over the total arc length yields the law of conservation of the area of the droplet, ${\cal A} $:
\be
{\cal A} \, =\, {1\over 2} \oint dl  \, \hat{\bf n} \cdot {\bf r} \, ,
\label{area}
\ee
\be
{d {\cal A}\over dt} \, =\, \oint dl \, \hat{\bf n}\cdot \dot{\bf r} \, \propto \,   \oint dl \, \partial_l j \, \equiv  0  \, \, .
\label{cons_are}
\ee


\subsection{Lagrangian}

The total energy of the system,
\be
E_{ \rm  tot}[\phi, {\bf r}]\, =\, E_{ \rm  SF}[\phi, {\bf r}] \, +\, E_{ \rm  cnf}[{\bf r}] \, ,
\label{E_tot}
\ee
splits into two distinct parts: the kinetic energy of superfluid currents,
\be
E_{ \rm  SF}[\phi, {\bf r}]\, =\,{n_s \over 2}  \! \oint \! dl (\partial_l \phi) ^2 \, ,
\label{E_SF}
\ee
and the configurational energy, which, in a general case of external potential,  includes two terms:
\be
E_{ \rm  cnf}[{\bf r}] \, =\, \chi \oint \! dl \, +\, E_{\rm pot}[{\bf r}] \, .
\label{E_cnf}
\ee
The first term is proportional to the total arc length. As it was done previously with $n_s$,
we ignore the effects of curvature, anisotropy, etc., and treat $\chi$ as a constant.
The second term is the potential energy of the droplet in an external potential, $U({\bf r}')$:
\be
E_{\rm pot}\, =\, \int_{\cal A} \!  U({\bf r}') \, d^2 r'
\label{E_pot_gen}
\ee
(the integration is over the position  ${\bf r}'$ inside the droplet area).

To produce the Lagrangian, we need to combine the energy with the term conjugating superfluid phase field to the edge shape degrees of freedom. The structure of this term can be guessed based on the special
case of an almost straight edge with the parameterization $\phi \equiv \phi(x)$, $y \equiv  y(x)$ considered in Ref.~\cite{sclimb}. In this case, the conjugating term has the form $\int d x \phi \dot y$  suggesting a straightforward generalization to
\be
 \! \oint \! dl \, \phi \, \hat{\bf n} \! \cdot \! \dot{\bf r}  = \int \!  \phi (y_\xi \dot{x} - x_\xi \dot{y})\, d\xi \, ,
\label{first_term}
\ee
thus leading to the Lagrangian 
\be
{\cal L}\, =\,  \! \oint \! dl \, \phi \, \hat{\bf n} \! \cdot \! \dot{\bf r} \, -\, E_{ \rm  tot}[\phi, {\bf r}] \, .
\label{L_gen}
\ee
The validation of the correctness of the form of the first term of Lagrangian (\ref{L_gen}) comes from the fact that variation over $\phi$ correctly reproduces continuity equation (\ref{cont}).

\subsection{Central potential}

Of a particular interest is the case of central potential $U({\bf r})=U(r)$, where polar parameterization (\ref{polar_x})--(\ref{polar_y}) with $x_0(t)=y_0(t)=0$ is the most natural. In this case,
Lagrangian (\ref{L_gen}) takes the form
\be
{\cal L}=- \oint d\theta  \left[ { \dot{\phi} r^2 \over 2}+ \frac{n_s \phi_\theta^2}{2\sqrt{r_\theta^2 + r^2}} +\chi \sqrt{r_\theta^2 + r^2}  + \Lambda(r) \right]  ,
\label{L_polar}
\ee
\be
\Lambda(r) \, =\, \int_0^r U(r') r' dr' \, .
\label{Lambda}
\ee
Note that (as we already mentioned) the variable $r^2/2$ is canonically conjugate to $\phi$.

\subsection{Generalized Beliaev–Josephson–Anderson equation}

Observing that the variation  of the term  $E_{\rm pot}[{\bf r}]$  with respect to the edge position can be written as
\be
\delta E_{\rm pot}[{\bf r}] \, = \, \oint   (y_\xi \delta x - x_\xi \delta y) \,  U(x(\xi),y(\xi)) \, d\xi \, ,
\label{dU}
\ee
we then find that variations of (\ref{L_gen}) with respect to $x$ and $y$ yield, respectively, two equations:
\be
-\dot{\phi} y_\xi + \dot{y} \phi_\xi - \partial_\xi \left[\left(\frac{n_s \phi_\xi^2}{2l_\xi^3} - \frac{\chi}{l_\xi}\right) x_\xi\right] - y_\xi U=0 \, ,
\label{E1}
\ee
\be
\dot{\phi} x_\xi - \dot{x} \phi_\xi -  \partial_\xi \left[\left(\frac{n_s \phi_\xi^2}{2l_\xi^3} - \frac{\chi}{l_\xi}\right) y_\xi\right] + x_\xi U=0 \, .
\label{E2}
\ee
Multiplying (\ref{E1}) by $x_\xi$ and (\ref{E2}) by $y_\xi$, and then adding the results
produces the anticipated identity, if the continuity equation (\ref{cont}) is taken into account.
Multiplying (\ref{E1}) by $y_\xi$ and (\ref{E2}) by $x_\xi$ and then subtracting the results, we obtain the the BJA relation as
\be
\dot{\phi} - (\hat{\bf t} \! \cdot\! \dot{\bf r}) \,   \partial_l \phi  \, = \,   \kappa({\bf r})\left [ {n_s\over 2 }(\partial_l \phi) ^2  - \chi  \right]  - U({\bf r}) \, .
\label{BJA_gen}
\ee
Note the covariant form of this equation.

\section{Non-Galilean ballistics}
Here we are interested in the motion of the droplet as a whole, that is, without changing its shape in time.
 In what follows, we will consider the case $\dot{r}=0$ in the parametrization (\ref{polar_x})--(\ref{polar_y}),  and use the reference frame of the stationary lattice unless otherwise specified. With the equations of motion (\ref{cont}) and  (\ref{BJA_gen}) we can readily prove (by contradiction) the absence of Galilean ballistics of the droplet. As a by-product, we will establish that the droplet will perform a uniform motion while having a circular shape only in the presence of the fine-tuned
external potential in the transverse to the motion direction.

Suppose we have a circular droplet of radius $r=R=$const moving along the $x$-axis  with the velocity $v_0$. In Eqs.~(\ref{polar_y}) we then have $y_0=0, x_0=v_0t$, that is, $\hat{\bf n}\!\cdot\! \dot{\bf r}=v_0 \cos \theta$  and  $\partial_l \, =\, R^{-1} \partial \theta$, so that Eq.~(\ref{cont}) implies
\be
\phi(\theta,t) \, =\, {v_0 R^2 \over n_s} \cos \theta + M\theta -\mu t \, ,
\label{phi_equilibr}
\ee
where $M$ is the phase winding number and $\mu$ is a certain constant.
If the lattice frame of reference---the one we work in---is inertial, then $M$ is an integer.
If the lattice rotates with angular velocity $\Omega$, then in its reference frame integer values of $M$ have to be shifted by $\Delta M = \varphi_0/(2\pi) = m_0\Omega R^2$, where $m_0$ is the particle mass and $\varphi_0 \in [-\pi, \pi]$ is the rotation-induced phase shift; here and in what follows we set $\hbar=1$.
The dependence on $\varphi_0$ adds a gyrometric aspect to the problem.
Linearity of the time-dependent additive term in (\ref{phi_equilibr}) is required by consistency with the time-independent r.h.s. of Eq.~(\ref{BJA_gen}).

Substitution of $\phi(\theta,t)$ of Eq.~(\ref{phi_equilibr}) into Eq.~(\ref{BJA_gen}) shows that there should be a fine-tuned external potential
\be
U(y) \, =\,  {3v_0^2 y^2 \over 2n_s R} \, - \,  {2Mv_0 y \over R^2} \, ,
\label{U_y}
\ee
and  the constant has been dropped.

The $M=0$ result reveals inconsistency between the uniform motion and circular droplet shape
in the absence of the external potential. To satisfy Eq.~(\ref{BJA_gen}) at low velocity,
\be
v_0 \, \ll \, {\sqrt{n_s\chi} \over R}\, ,
\label{circ_cond}
\ee
one has to assume small deformation of the droplet shape (and thus its curvature),
$r(\theta ) = R - \epsilon f(\theta )$. Keeping only linear in $\epsilon$ terms
in (\ref{BJA_gen}) we obtain
\be
\epsilon = \frac{v_0^2R^3 }{4 n_s \chi}  \, \ll\, R \,, \qquad f=\cos (2\theta) \,,
\label{shape}
\ee
 {\it i.e.}, in the absence of external potential, the distortion of the circular shape
(elongation in the $y$-direction) of the moving droplet (with $M=0$) vanishes
quadratically with $v_0$ at $v_0\to 0$. The parabolic confining potential in (\ref{U_y})
is required to ``compress" the droplet back to its circular shape. From the flow kinetic energy,
$E_{\rm SF} = \pi R^3 v_0^2/2n_s$,  we also obtain the droplet effective mass as
\be
m_{\rm eff} = \frac{\pi R^3 }{n_s} +  {\cal O}(\epsilon^2) \, .
\label{meff}
\ee
The same result follows from the analysis of the sloshing mode considered in Sec.~\ref{modes}.
When the condition (\ref{circ_cond}) is violated, the superflow at the edge is no longer protected
by small parameters against either quantum phase slips or dynamic spectrum instability \cite{Radzihovsky2023};  also, at such values of 
$v_0$, the bi-linear in $\partial_l \phi$ form of $E_{\rm SF}$ cannot be justified. 

The $M \ne 0$ case is fundamentally different. To begin with, in the absence of external potential
it is impossible to satisfy Eq.~(\ref{BJA_gen}) by deforming the droplet shape: Mathematically,
the problem reduces to solving equation
\[
f+f_{\theta \theta} =-3\cos (2\theta ) +b\sin \theta \, ,
\]
which has no $2\pi$-periodic solutions when $b \propto M$ is non-zero.
In other words, persistent current at the droplet edge eliminates
the possibility of the ballistic propagation in free space!  Uniform motion at the velocity $v_0$ along $x$ becomes possible, if both the droplet deformation
\be
r= R- \epsilon \cos(2\theta)
\label{def_F}
\ee
occurs and the uniform force
\be
F=\frac{2Mv_0}{R^2},
\label{Fy}
\ee
with $U=-Fy$  along $y$ is applied. [It is assumed that $M^2n_s/(2 \chi R^2) \ll 1$]. It is worth mentioning that such linear potential can be induced by accelerating the whole lattice.

Of special interest is the case $U=0, \, M\neq 0$. As shown above, no uniform motion is possible in this case. The solution to the system of equations (\ref{cont}) and (\ref{BJA_gen})
exits when the droplet performs a centripetal motion with some radius $R_c$ at some angular velocity $\omega_c$ and simultaneously is deformed.
We consider the case of small deformation $r=R- \epsilon f(\theta)$.
Then, we find $\hat{\bf n}\! \cdot \! {\bf r}=R_c \omega_c \sin(\theta - \omega_ct)$ and $\hat{\bf t}\! \cdot \! {\bf r}= R_c \omega_c \cos( \theta -\omega_ct)$. The continuity equation
gives $\phi= (R_c \omega_c R^2/n_s) \sin(\omega -\omega_ct) +M \theta -\mu t$. Substituting $\phi$ into Eq.~(\ref{BJA_gen}) gives
\be
r=R- \frac{(R_c \omega_c)^2 R^3\cos[2(\theta - \omega_c t)] }{4n_s\chi } + {\cal O}(\epsilon^2)
\label{cen}
\ee
and
\be
\omega_c =\frac{2Mn_s}{R^3}.
\label{omc}
\ee
This situation resembles centripetal motion of a particle with mass (\ref{meff}) carrying some charge $q$ in the magnetic field $B=2\pi M/q$.

\section{Ground-state solution and AC Josephson effect}

As in any superfluid, the ground state of the droplet features broken time-translation symmetry---the time-crystallization effect---manifested by the linear growth of the phase with time:
\be
\dot{\phi} \, =\, - \mu \, ,
\label{mu}
\ee
where $\mu = dE/dN$ is the chemical potential that depends on the area and shape of the droplet.
The latter is sensitive to the presence of anisotropic trapping potential, in which case the shape becomes also sensitive to the presence of the supercurrent.
The continuity equation states that the phase gradient along the edge is constant:
\be
\partial_l \phi \, =\, \zeta \, .
\label{cont_stat}
\ee
The two parameters, $\mu$ and $\zeta$, control the shape of the droplet via the stationary BJA equation:
\be
\left( \chi- {n_s\zeta^2 \over 2} \right) \kappa({\bf r}) + U({\bf r})\, =\, \mu \, .
\label{BJA_stat}
\ee
In the general case of an anisotropic potential $U({\bf r})$ and nonzero supercurrent,
the parameter $\zeta$ can be viewed at the eigenvalue of the problem at a given value of $\mu$. It has to satisfy the phase-winding quantization condition
\be
\zeta \oint dl \, =\, 2\pi M \, .
\label{quantization}
\ee
 (Equivalently, $\mu$ can be viewed as an eigenvalue of the problem at a given value $\zeta$.) For a given phase winding $M$,
we thus get a single-parametric family of solutions controlled by the pair $(\mu, \zeta_M(\mu))$ that implicitly defines the shape of the droplet as a function of the total amount of matter and $M$.

In the case of isotropic potential, the situation is very simple. The droplet has a circular form, meaning that $\kappa = 1/R$ and $\zeta = M/R$, where
$R$ is the radius of the droplet. Equation (\ref{BJA_stat}) then simply relates $\mu$ to $R$ and $M$:
\be
\mu \, =\, U(r) + {1\over R}\left( \chi- {n_s M^2 \over 2 R^2} \right) \, .
\label{BJA_circle}
\ee

\section{Normal modes.\\ Effect of a supercurrent}
\label{modes}

To find normal modes of a circular droplet of radius $R$ trapped in a rotationally symmetric potential we
need to linearize equations of motion in the vicinity of the equilibrium solution
with $\mu$ given by Eq.~(\ref{BJA_circle}). This is done by substituting
\bea
\phi(\theta, t)  \, &=&\, -\mu t + M \theta + \varphi(\theta, t)  \, ,
\label{phi_lin} \\
 r(\theta, t) \,& =& \, R + h(\theta, t) \, ,
\label{r_lin}
\eea
either into equations of motion (\ref{cont}) and (\ref{BJA_gen})
or directly into the Lagrangian (\ref{L_polar}).
In the latter case---implemented below, the linear in $\varphi$ and $h$ terms automatically nullify and the resulting bi-linear Lagrangian generates the desired pair of linear in $\varphi$ and $h$ dynamic equations describing the normal modes, including the ballistic motion in the absence of the trapping potential.

The bi-linear Lagrangian reads
\be
{\cal L}_{\rm bl}=- \! R\oint \! d\theta  \left[ \dot{\varphi}h+ {A\varphi_\theta^2\over 2} -B\varphi_\theta h + {C h_\theta^2  \over 2} + {D h^2  \over 2} \right]  ,
\label{L_pol_bilin}
\ee
\be
A\, =\, {n_s\over R^2} \, , \qquad B \, =\, {n_sM\over R^3} \, ,
\label{A_B}
\ee
\be
C = {\chi\over R^2} -  {n_s M^2\over 2R^4} \, , \quad D  = U'(R) - {\chi\over R^2} +  {3 n_s M^2\over 2R^4}  .
\label{C_D}
\ee
The solution to the equations of motion,
\be
\dot{h} + A\varphi_{\theta \theta} -Bh_\theta = 0 \, ,
\label{EoM_h}
\ee
\be
\dot{\varphi} -B\varphi_\theta - Ch_{\theta \theta} + D h = 0 \, ,
\label{EoM_phi}
\ee
is a linear combination of normal modes
\be
h_m= {\rm Re}\, \alpha_m e^{im\theta - i\omega_m^{\pm} t} , \quad \varphi_m ={\rm Re}\, \beta_m e^{im\theta - i\omega_m^{\pm}  t} \,
\label{norm_modes}
\ee
\be
\omega_m^{\pm}  = Bm \, \pm\,  m \sqrt{A(m^2C + D)}  , \quad m=1,2,3, \ldots \, .
\label{omega_m_gen}
\ee

With Eqs.~(\ref{A_B})--(\ref{C_D}) we have
\be
\omega_m^{\pm}  = {m n_s M \over R^3} \pm   m \sqrt{ {n_sU'(R)\over R^2}  + {n_s^2 M^2 \over R^6} + {(m^2\! -\! 1) n_s \tilde{\chi} \over R^4} }\,  ,
\label{omega_m}
\ee
where
\be
\tilde{\chi} \, =\, \chi - {n_s M^2 \over 2R^2}
\label{tilde_chi}
\ee
is a  renormalized (due to the supercurrent) parameter $\chi$. In what follows, we assume
that this renormalization is small because the condition
\be
|M| \, \ll\, \sqrt{\chi/n_s} \, R \, ,
\label{tilde_chi_small}
\ee
protects supercurrent states from quantum phase slips.

\subsection{Fundamental Modes}

Of special interest are fundamental solutions corresponding to $m=1$.
Their frequencies do not depend on $\chi$ but do depend on $M$:
\be
\omega_1^{\pm}  = {n_s M \over R^3}\, \pm\,   \sqrt{{n_s\over R^2}\left[ U'(R) + {n_s M^2 \over R^4} \right]} .
\label{omega_1}
\ee
In the absence of supercurrents, both frequencies are equal (up to the global sign) and the fundamental solution corresponds to the doubly degenerate sloshing mode with frequency
\be
\omega_{\rm sl}  =  {\sqrt{n_s \, U'(R) } \over R} \, .
\label{sloshing}
\ee
In the limit of $U'(R)\to 0$, this mode corresponds to ballistic motion with near-circular droplet shape;
{\it i.e.},  it is identical to that of a point particle
with effective mass (\ref{meff}) in the potential $ \pi R^2 U(r)$.

Supercurrent qualitatively changes the picture of motion. The two frequencies become
different leading to two characteristic regimes controlled by the value of the parameter
\be
\gamma \, =\,   {|M| \over R^2 } \sqrt{n_s  \over U'(R)} \, .
\label{gamma}
\ee
At $\gamma \ll 1$,  the relative difference between the magnitudes of the two frequencies is small:
\be
|\omega_1^{\pm}| \, =\,   \omega_{\rm sl} \left(\sqrt{1+\gamma^2} \, \pm\,  \gamma \right) .
\label{gamma_ll_1}
\ee
Here we are dealing with the previously-discussed sloshing mode that now demonstrates
slow Foucault-type precession.

In the regime $\gamma \gg 1$, we have $|\omega_1^{-}| \ll |\omega_1^{+} |$:
\be
|\omega_1^{\pm}| \, =\, \omega_* {\sqrt{1+\gamma^{-2}} \, \pm\,  1 \over 2} , \qquad  \omega_*  =   {2 n_s |M| \over R^3} \, .
\label{gamma_gg_1}
\ee
Here the motion is similar to that of a 2D charged particle in perpendicular to the plane magnetic field and weak harmonic trap. In particular, this means that there is no ballistic motion at $M\neq 0$. Indeed, in the absence of the external potential, $|\omega_1^{+} |= \omega_*$ and $\omega_1^{-}=0$, implying that the center of mass
performs uniform circular motion with the angular frequency $\omega_*$.

\subsection{Modes with $m\gg 1$}
At $m\gg1$ we can neglect the middle term under the square root in Eq.~(\ref{omega_m}) because inequality (\ref{tilde_chi_small}) guarantees that it is small and is getting progressively less relevant
with increasing $m$, and omit $1$ compared to $m^2$:
\be
|\omega_m^{\pm} | \to   m \sqrt{ {n_s\over R^2} \left[ U'(R) +  {m^2 \tilde{\chi} \over R^2}  \right] } \pm  {m n_s M \over R^3}  \quad (m\gg 1)\, .
\label{large_m}
\ee
For the same reasons the second term is a small correction.
If the external potential $U$ is appropriately weak or absent, we can also omit the term $U'(R)$:
\be
|\omega_m^{\pm}|  \to    {m^2 \sqrt{n_s \tilde{\chi} }\over R^2} \pm {m n_s M \over R^3}  \qquad \left( {R^2 \, U'(R)\over \chi} \ll m^2 \right) ,
\label{large_m_2}
\ee
to recover the quadratic dispersion of elementary excitations in the TQF state of a straight edge \cite{Kuklov2022}.
At $U'(R) \gg \chi/R^2 $, there emerges a range of $m$ values where the dispersion is linear in $m$
and independent of $\chi$, reflecting the fact that the potential of that strength converts the edge
into a Luttinger liquid.

\section{Conclusions and Outlook}
\label{sec:conclus}

A two-dimensional insulating domain (``droplet") with a superclimbing edge, see Fig.~\ref{fig1},
can be formed in a system of hard-core bosons with nearest-neighbor attraction tuned to guarantee, on the one hand, a phase-separated ground state, and, on the other hand, wide enough---and thus microscopically quantum-rough---edge.
The counterintuitive autonomous dynamics of such a domain is controlled by and is characteristic
of the most unusual properties of Transverse Quantum Fluid (TQF) formed at the droplet edge.

The  supertransport along the edge enables coherent (dissipation-free) displacement of the edge---the superclimbing motion. For an isolated droplet, as opposed to an infinitely long edge, or an edge with pinned ends, the supeclimbing motion features fundamental modes sensitive to the presence of circulating supercurrent along the edge. In the translation invariant case and in the absence of circulating current, the droplet moves pseudo-ballistically while preserving its
near circular shape---apart from slight, proportional to the square of the velocity, elongation in the transverse to the displacement  direction. ``Pseudo" refers to the fact that the bulk currents are zero: the insulating domain propagates in space exclusively through the matter transfer by edge supercurrents. An isotropic trapping potential converts the pseudo-ballistic motion into the sloshing mode.

The circulating supercurrent along the edge dramatically changes the droplet dynamics: The motion acquires features resembling that of a gyroscope or a 2D charged particle in a perpendicular magnetic field. In a linear external potential (uniform force field), a droplet with a circulating supercurrent demonstrates a spectacular gyroscopic effect---uniform motion in the perpendicular to the force direction. This effect has a natural gyrometric aspect
when the lattice rotates; the rotation-induced phase twist results in finite superrcurrent circulation
in the reference frame of the lattice.

As in any superfluid, the ground state of the droplet features broken time-translation symmetry---the time-crystallization effect---manifested in the linear growth of the phase with time, $\dot{\phi}=-\mu t$. The period, $2\pi/\mu$,  of the superfluid phase evolution in the ground-state (dictating the frequency of the AC Josephson effect) is sensitive to the size (as well as other geometric details) of the droplet; see  Eq.~(\ref{BJA_circle}) for the case of a circle.

On the technical side, dynamics of the droplet is described by Lagrangian formalism in terms of the edge position, ${\bf r}(\xi,t)=(x(\xi,t),y(\xi,t))$, and the superfluid phase along the edge, $\phi(\xi,t)$, as functions of time  $t$, with parameter $\xi$ labelling the edge points. The structure of the Lagrangian, Eq.~(\ref{L_gen}), readily follows from the continuity equation, the form of the latter expressing the law of conservation of matter under specific conditions of (i) supertransport along the edge and (ii) insulating incompressible bulk. The two Euler-Lagrange equations implied by the Lagrangian (\ref{L_gen}) are (i) the continuity equation (\ref{cont}) and (ii)  the generalized Beliaev--Josephson--Anderson equation (\ref{BJA_gen}).

Numerous other possible physical implementations of the autonomous superclimbing droplet include multi-component
bosons and higher-spin $XY$-magnets, fermionic rather than bosonic systems of ultracold atoms,
\he4 and/or \hel3 domains on substrates or complete layers of similar atoms, superclimbing edge dislocation
loops in \he4 and/or \hel3. Finally, similar phenomena may takes place in 3D with an insulating ball having a superclimbing surface.

\begin{acknowledgements}
We acknowledge support from the National Science Foundation under Grants DMR-2335905 and DMR-2335904.
\end{acknowledgements}

\end{document}